# Using VGG16 Algorithms for classification of lung cancer in CT scans Image

**Hasan Hejbari Zargar [1], Saha Hejbari Zargar [2, 3], Raziye Mehri [2, 3], Farzane Tajdini [4]**

1. Islamic Azad University of Ardabil, Ardabil, Iran
2. Deputy of Research and Technology, Ardabil University of Medical Sciences, Ardabil, Iran
3. Department of Community Medicine, Faculty of Medicine, Ardabil University of Medical Science, Ardabil, Iran
4. Tabarestan University of Chalus, Chalus, Iran

## Abstract

Lung cancer is the leading reason behind cancer-related deaths within the world. Early detection of lung nodules is vital for increasing the survival rate of cancer patients. Traditionally, physicians should manually identify the world suspected of getting carcinoma. When developing these detection systems, the arbitrariness of lung nodules' shape, size, and texture could be a challenge. Many studies showed the applied of computer vision algorithms to accurate diagnosis and classification of lung nodules. A deep learning algorithm called the VGG16 was developed during this paper to help medical professionals diagnose and classify carcinoma nodules. VGG16 can classify medical images of carcinoma in malignant, benign, and healthy patients. This paper showed that nodule detection using this single neural network had 92.08% sensitivity, 91% accuracy, and an AUC of 93%.

# 1. Introduction

Lung cancer is a complex disease composed of diverse histological and molecular types with clinical relevance. The advent of large-scale molecular profiling has been helpful to identify novel molecular targets that can be applied to the treatment of particular lung cancer patients and has helped to reshape the pathological classification of lung cancer. Novel directions include the immunotherapy revolution, which has opened the door for new opportunities for cancer therapy and is also redefining the classification of multiple tumors, including lung cancer [1].

In recent years, where computers try to learn or discover the hidden pattern of educational data, the use of machine learning technology has become increasingly popular [2]. However, analyzing slides manufactured from the biopsy could be a difficult task for a pathologist and requires considerable time, skill, and considerable tact. This can not be error-free and should not detect small metastases. Compared to this, computer-aided histopathological analysis can play an essential role in diagnosing and prognosis of BC tumor cells. In this case, AI (AI) algorithms are an endeavor to enhance the speed and accuracy of detection [3].

Medical image analysis has extraordinary supremacy in the field of health sector, particularly in noninvasive treatment and clinical examination. The acquired restorative images such as X-rays, CT, MRI, and ultrasound imaging are used for specific diagnosis. In medical imaging, CT is one of the filtering mechanisms which use attractive fields to capture images in films. Lung cancer is one-of-its-kind of cancer that leads to 1.61 million deaths per year. The survival rate is higher if the cancer is diagnosed at the beginning stages. The early discovery of lung cancer is not a simple assignment. Around 80% of the patients are diagnosed effectively only at the center or propelled phase of cancer. Early detection of cancer is the most promising way to enhance a patient's chance for survival. This paper presents a computer aided classification method in computed tomography (CT) images of lungs developed using artificial neural network. The classification process is done by VGG16 [4].

# 2. Related Work

Lung cancer is positioned second among males and tenth among females globally. The information given in these studies is a general portrayal of lung cancer location framework that contains four basic stages. The lung cancer is the third most frequent cancer in women, after breast and colorectal cancers. Feature extraction process is one of the simplest and efficient dimensionality reduction techniques in image processing. One of the striking features of CT imaging is its non-obtrusive character. The rise of angles, which might be viewed, is odd when compared to parallel imaging modalities [4].

The carcinoma diagnosis has previously been studied using image processing techniques [5-7]. With the arrival of neural networks and deep learning techniques, these cases have recently been utilized in the sector of medical imaging [8-10]. Various researchers [11-15] have tried to classify and diagnose carcinoma using machine learning and neural networks. Many in-depth learning methods have not been wonted to diagnose carcinoma. This is often thanks to the shortage of an in-depth data set for medical images, especially carcinoma. Shimizu et al. [16, 17] Use urine samples to diagnose carcinoma.

## 3. MATERIAL AND METHODS
### 3.1. Dataset
The data set is available at kaggle, (IQ-OTH/NCCD - Lung Cancer Dataset). Finally, the system is evaluated using Python-Jupyter Notebook-based simulation results. 80% of the data is for training, and 20% of the information is for testing.

### 3.2. Pre-processing
Preprocessing refers to raw data evolution before it is transferred to a machine learning or deep learning algorithm. For example, training a convolutional neural network in raw images will likely lead to poorly classified functions. Pre-processing is also important to speed up learning. In the preprocessing step, we prepare the data for feeding to the Keras model.

We also split the data into a training and testing suite. Finally, we will scale the data and standardize it from one to one. This standard helps to model the model better and makes it more convergent.

### 3.3. Evaluation metrics
The confusion matrix is the N X N matrix, where N is the number of categories predicted. For the prevailing problem, we've N = 2 and thus obtain a 2 X 2 matrix. Here are some definitions you ought to detain mind for a confusion matrix:

**Accuracy**: Indicate the class's number of "correct predictions made" divided by the number of "total predictions made" by the same category [39, 40].

$$Accuracy = \frac{TP + TN}{TP + FP + TN + FN} \quad (1)$$

**Sensitivity**: Real positive rate: If the result is positive for the person, in a few percent of cases, the model will be positive, which is calculated from the following formula.

$$Sensitivity = \frac{TP}{TP + FN} \quad (2)$$

**Properties**: Real negative rate: If the result is negative for the person, in a few percent of cases, the model will also be a negative result, which is calculated from the following formula.

$$Specificity = \frac{TN}{TN+FP} \quad (3)$$

**Recall**: The recall criterion expresses the ratio of the "number of correctly categorized text data" in a particular class to the total number of data that must be categorized for that specific class.

$$Recall = \frac{TP}{TP+FN} \quad (4)$$

**Precision** measures the ratio of the division of "correctly made predictions" for samples of a particular class to the number of "total predictions" for examples of the same class (this number includes the sum of all true and false predictions) [39, 40].

$$\text{Precision} = \frac{TP}{TP+FP} \qquad (5)$$

**TABLE 1. Confusion Matrix**

| Actual class / Predicted class | Cancer | Normal |
|---|---|---|
| **Cancer** | true positives | false positive |
| **Normal** | false negative | true negatives |

## 4. RESULTS

Medical imaging data has shown an explosive growth trend, and AI and deep learning technology are evolving. Engineering has become more precise and intelligent. Within the context of in-depth medical data, using deep learning technology to comprehend a more efficient computer-assisted system has become a brand-new challenge within the medical field. Effective diagnosis of pulmonary nodules plays a vital role in the early detection of lung cancer. Accordingly, researchers have done much research. Within the early stages of pulmonary node diagnosis, researchers mainly use traditional machine learning methods to diagnose pulmonary nodules [18]. In short, a computer-assisted and computer-assisted lung node detection system is crucial for the first detection of carcinoma.

As shown in Table 2 and Figure 1, the accuracy of lung cancer diagnosis in different algorithms was compared with the validation approach in terms of rate.

**Table 2. Results of algorithms CNN**

| Model | VGG16 | | | |
|---|---|---|---|---|
| | Approach | Accuracy | Sensitivity | Specific |
| VGG16 | Hold out | 89.08 | 90.00 | 88.12 |
| | K-fold=5 | 90.25 | 88.55 | 88.98 |
| | K-fold=10 | 92.08 | 91.00 | 93.01 |

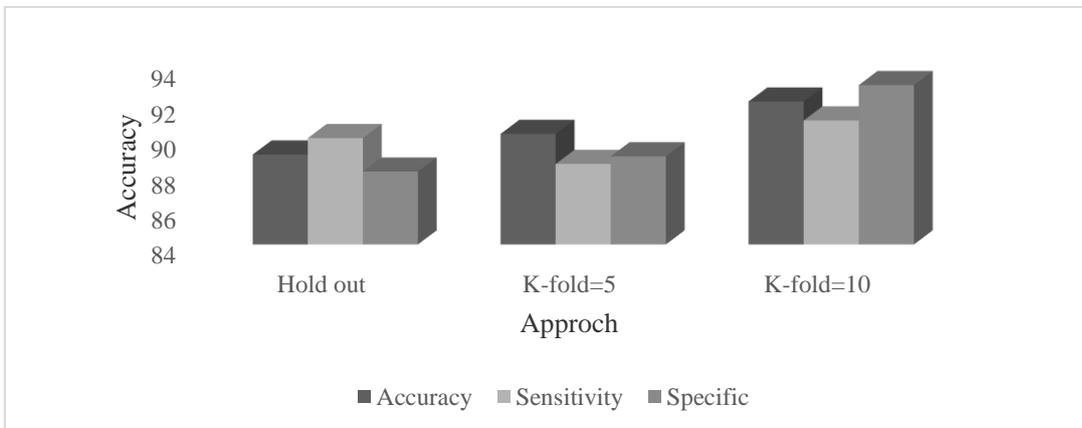

**Fig 1. Results of algorithms CNN**

5. Conclusion

Machine learning and deep learning [28-37] has been successfully utilized in medical image and healthcare [2] analysis like covid-19 [19], whole-slide pathology images [20], Types of the Epidemic from X-rays [21], detection of breast cancer metastasis [22], Identification of Medicinal Plants [23], Investigation [24], predict the time between symptom onset and hospital arrival in stroke patients and its related factors [25], Maximizing the Impact on Social Networks [26], outcome prediction of bupropion exposure [27], etc. Lung cancer is common and has a high mortality and morbidity. Outcome is dependent on clinical stage and cancer cell type. Using ctscan may be classified according to the vgg16 algorithm.